\documentclass[11pt,a4paper]{article}
\pdfoutput=1
\usepackage{jcappub}
\usepackage{pdflscape}
\newcommand{\mr}[1]{\mathrm{#1}}
\newcommand{\wde}{w_{\mr{de}}}
\newcommand{\rhoc}{\rho_{\mr{c}}}
\newcommand{\rhode}{\rho_{\mr{de}}}
\newcommand{\h}{\mathcal{H}}
\newcommand{\aref}{a_{\mr{ref}}}
\newcommand{\tauref}{\tau_{\mr{ref}}}
\newcommand{\rref}{r_{\mr{ref}}}

\title{Distinguishing interacting dark energy from $w$CDM with CMB, lensing, and baryon acoustic oscillation data}

\author{Jussi V\"aliviita}
\author{and Elina Palmgren}
\affiliation{Department of Physics and Helsinki Institute of Physics, University of Helsinki,\\ P.O. Box 64, FIN-00014 University of Helsinki, Finland}

\emailAdd{jussi.valiviita@helsinki.fi}
\emailAdd{elina.palmgren@helsinki.fi}

\abstract{We employ the Planck 2013 CMB temperature anisotropy and lensing data, and baryon acoustic oscillation (BAO) data to constrain a phenomenological $w$CDM model, where dark matter and dark energy interact. We assume time-dependent equation of state parameter for dark energy, and treat dark matter and dark energy as fluids whose energy-exchange rate is proportional to the dark-matter density.  The CMB data alone leave a strong degeneracy between the interaction rate and the physical CDM density parameter today, $\omega_c$, allowing a large interaction rate $|\Gamma| \sim H_0$. However, as has been known for a while, the BAO data break this degeneracy. Moreover, we exploit the CMB lensing potential likelihood, which probes the matter perturbations at redshift $z \sim 2$ and is very sensitive to the growth of structure, and hence one of the tools for discerning between the $\Lambda$CDM model and its alternatives. However, we find that in the non-phantom models ($w_{\rm de}>-1$), the constraints remain unchanged by the inclusion of the lensing data and consistent with zero interaction, $-0.14 < \Gamma/H_0 < 0.02$ at 95\% CL. On the contrary, in the phantom models
($w_{\rm de}<-1$), energy transfer from dark energy to dark matter is moderately favoured over the non-interacting model; $-0.57 < \Gamma/H_0 < -0.10$  at 95\% CL with CMB+BAO, while addition of the lensing data shifts this to $-0.46 < \Gamma/H_0 < -0.01$.}

\keywords{cosmological parameters from CMBR, cosmological parameters from LSS, dark energy theory, dark matter theory}

\arxivnumber{1504.02464}

\newcommand{\jvcomment}[1]{}

\begin{document}

\begin{flushleft}
	\hfill			HIP-2015-11/TH
\end{flushleft}

\maketitle

\section{Introduction}

The recently published Planck full-sky maps of the anisotropies of the cosmic microwave background (CMB) radiation are, in general, compatible with a simple 6-parameter cosmological ``standard'' model \cite{Ade:2013sjv,Ade:2013kta,Ade:2013zuv,Planck:2013jfk,Planck:2013nga,Ade:2015ktc,Ade:2015zuv,Ade:2015uln},  the $\Lambda$CDM model, which was also in a good agreement with the earlier, less accurate, measurements. Apart from the 5\% ordinary matter and radiation components, two exotic building blocks are needed. 68\% of the energy density of the Universe comes from dark energy and 27\% from dark matter. Dark energy causes the accelerating expansion of the Universe and dark matter is an invisible --- weakly interacting or noninteracting --- component that is needed to explain the galaxy rotation curves~\cite{Sofue:2000jx}, the structure formation, and the flatness of the spatial geometry of the Universe.
Indeed, cosmological models with only ordinary matter do not provide a good fit to a broad variety of cosmological and astrophysical data. Thus, it can be regarded well established that some kind of ``dark sector'' is needed.

Although the micro-physics of the dark sector is still largely unknown, many of its properties can be tested or constrained by comparing theoretical predictions to the CMB and large scale structure (LSS) observations. It has been proposed that the origin of both dark matter and dark energy could be some kind of scalar fields. From a particle physics point of view, it would be natural to assume that these fields interact with each other or with (dark) matter \cite{Wetterich:1994bg,Amendola:1999qq,Amendola:1999dr,Holden:1999hm,Billyard:2000bh,Hwang:2001fb,Zimdahl:2001ar,TocchiniValentini:2001ty,Comelli:2003cv,Chimento:2003iea,Farrar:2003uw,Amendola:2003wa,Manera:2005ct,Koivisto:2005nr,Das:2005yj,Amendola:2006qi,Bean:2007ny,Bean:2008ac,Chongchitnan:2008ry,Corasaniti:2008kx,Chen:2008ft,LaVacca:2009yp,Li:2009sy,Xia:2009zzb,deSouza:2010ym,LopezHonorez:2010ij,Li:2010eu,Beyer:2010mt,Tarrant:2011qe,Pavan:2011xn,Ziaeepour:2011bq,Aviles:2011ak,Pettorino:2012ts,Pourtsidou:2013nha}. The dark matter could ``decay'' to dark energy and vice versa. However, in the cosmological standard model, dark matter and dark energy are assumed to feel only each others gravitational effects. In this paper we relax this hypothesis by allowing for a phenomenological energy transfer term, i.e., interaction between dark matter and dark energy. 

We treat the dark matter  ($\mr{c}$) and dark energy ($\mr{de}$) as fluids that have equation of state parameters $w_{\rm{c}}=0$ and $\wde(t)=p_{\mr{de}}(t)/\rhode(t)$ and whose energy-exchange rate is proportional to the dark matter density. While the total energy density obeys the usual continuity equation,
\begin{equation}
  \rho_{\mr{tot}}' + 3 \h (1+w_{\mr{tot}}) \rho_{\mr{tot}} = 0,
\end{equation}
the model is characterized in the background by modified equations for the individual components,
\begin{eqnarray}
\rhoc'  + 3 \h \rhoc & = & (a/\aref) Q_{\rm{c}}\,, \label{eq:cont_rhoc} \\
\rhode' + 3 \h (1+\wde) \rhode & = & (a/\aref) Q_{\mr{de}}\,,\label{eq:cont_rhode}
\end{eqnarray}
where $Q_{\mr{de}}=-Q_{\mr{c}}$, a prime indicates derivative with respect
to conformal time $\tau$, $a$ is the scale factor of the Universe, and $\h = a'/a$  is the conformal Hubble parameter. The scale factor at a reference time, $\aref$, is usually chosen to be today's scale factor and normalized to unity, $\aref = a_0 = 1$, but for later convenience we keep it in the above equations.

In the literature, there are three main categories for the choice of the phenomenological energy exchange term $Q_{\mr{c}}$:
\begin{itemize}
\item[(1)] For example in \cite{Olivares:2005tb,Olivares:2006jr,Sadjadi:2006qp,Amendola:2006dg,Guo:2007zk,Olivares:2007rt,Abdalla:2007rd,Boehmer:2008av,He:2008tn,Pereira:2008at,Quercellini:2008vh,He:2008si,CalderaCabral:2008bx,Gavela:2009cy,Jackson:2009mz,MohseniSadjadi:2009gs,He:2010ta,Karami:2010tr,Amirhashchi:2010pi,He:2010im,Xu:2011tsa,RozasFernandez:2011je,Jamil:2011iu,Shi:2012ma,Xu:2013jma,Wang:2014iua, Geng:2015ara, Duniya:2015nva} $Q_{\rm{c}}$ assumed to be proportional to the \emph{Hubble rate, $H$, times} either of the energy densities or their sum or some other combination of the energy densities.
In \cite{Amendola:2006dg} a phenomenological coupling of type $Q_{\rm{c}} = \mbox{constant} \times H \rhoc$ is motivated by a quintessence model which induces a time dependent mass for the dark matter particles. The energy transfer could also change its direction during the history of the Universe, as studied in \cite{Cai:2009ht} by fitting to the background data an interaction of the form of $\mbox{a piecewise constant} \times H\rho$. The above-mentioned papers either focus only on the background evolution or, if perturbations are included, assume that $H$ in the interaction term is spatially constant, i.e., assume it to be the average expansion rate.  However, when deriving the perturbation equations in a consistent way, one should treat $H$ as a local variable, i.e., include the term $\delta H$ \cite{Gavela:2010tm}, as also noticed in \cite{Martinelli:2010rt,DeBernardis:2011iw}.
\item[(2)] Another common choice for a phenomenological interaction is a constant times either of the energy densities or some combination of them (without including the Hubble parameter and hence an implicit time dependence), as done for example in \cite{Boehmer:2008av,Schaefer:2008ku,Quartin:2008px,Valiviita:2008iv,CalderaCabral:2008bx,CalderaCabral:2009ja,Koyama:2009gd,Majerotto:2009np,Valiviita:2009nu,Boehmer:2009tk,Koshelev:2009nj,Lip:2010dr,Tong:2011eb,Potter:2011nv,Clemson:2011an,Zhang:2013lea,Hashim:2014rda,Zhang:2013zyn,Geng:2015ara, Duniya:2015nva}. In other contexts, this type of interaction has been used to describe the decay of curvaton \cite{Malik:2002jb,Ferrer:2004nv,Sasaki:2006kq,Assadullahi:2007uw} or production of quintessence field condensation by a slow decay of superheavy dark matter \cite{Ziaeepour:2003qs}.
\item[(3)] An interesting possibility is also an elastic scattering between the dark-sector ``particles'', which does not lead to the energy exchange in the background. These models leave much milder observational signatures and indeed the current upper limits on the interaction cross-section in these models are several orders of magnitude larger than the Thomson cross-section $\sigma_T$ \cite{Simpson:2010vh}, if $\wde > -1$. On the other hand, tight constraints ($<10^{-9}\sigma_T$) are obtained if $\wde < -1$ \cite{Xu:2011nr}.
\end{itemize}
In this paper we will study a model where
\begin{equation}
Q_{\rm c} = -\Gamma \rho_{\mr{c}}\,,
\label{eq:ourQc}
\end{equation}
with $\Gamma=\,$constant. A positive $\Gamma$ describes an energy transfer or a decay of dark matter to dark energy ($\rhoc \rightarrow \rhode$ in the background) and a negative $\Gamma$ corresponds to an energy transfer from dark energy to dark matter ($\rhode \rightarrow \rhoc$ in the background). From equations \eqref{eq:cont_rhoc} and \eqref{eq:cont_rhode} we see that $\Gamma$ has the same unit as $H$. Therefore, it is convenient to describe the interaction by a dimensionless constant $\Gamma/H_0$, which gives the interaction rate in units of today's expansion rate. 

The interaction \eqref{eq:ourQc} falls into the category (2) and we will follow closely \cite{Valiviita:2008iv,Majerotto:2009np,Valiviita:2009nu}. Our first aim is to update the constraints\footnote{%
In \cite{Valiviita:2009nu} only models with $\wde > -1$ were constrained. Now we study also phantom models ($\wde < -1$) in separate Markov Chains Monte Carlo runs. 
} by using the Planck CMB temperature anisotropy data \cite{Ade:2013kta}, including the full treatment of perturbations and calculation of the CMB angular power spectrum. As seen, e.g., in \cite{Valiviita:2009nu}, the CMB data alone leave a strong degeneracy between the interaction rate and today's physical CDM density parameter $\omega_{c}$, since the CMB probes the CDM density at the time of last scattering at redshift $z_\ast\sim1100$, and the interaction modifies the evolution between last scattering surface and today. Therefore, it is important to choose a complementary data set that constrains the matter (or dark energy) density near to today. In \cite{Valiviita:2009nu} we showed that either baryon acoustic oscillation (BAO) data or supernova data are almost ``orthogonal'' to the CMB data when constraining the interacting dark-sector model; see also \cite{Lee:2009ji}. To break the degeneracies we choose now three recent BAO measurements that probe the redshift range $z_{\mr{eff}} \approx 0.1$ -- $0.6$: 6dF Galaxy Survey \cite{Beutler:2011hx} at $z_{\mr{eff}}=0.106$, Sloan Digital Sky Survey Data Release 7 (SDSS DR7) BAO \cite{Percival:2009xn} data point at $z_{\mr{eff}}=0.35$ as reanalyzed by \cite{Padmanabhan:2012hf}, and Baryon Oscillation Spectroscopic Survey Data Release 9 (BOSS DR9) \cite{Anderson:2012sa} at $z_{\mr{eff}}=0.57$. These choices coincide with those made in the Planck 2013 analysis in \cite{Ade:2013zuv}.\label{page:BAO}

Among many other measurements than just the primary CMB temperature anisotropy, Planck made a more than $25\sigma$ detection of weak gravitational lensing of CMB by large-scale structure \cite{Ade:2013tyw} (which in the 2015 release improved to 40$\sigma$ \cite{Ade:2015zua}), superseding the previous $\sim 5\sigma$ detections of Atacama Cosmology Telescope (ACT) \cite{Das:2013zf} and South Pole Telescope (SPT) \cite{vanEngelen:2012va}. The CMB lensing potential power spectrum, $C_\ell^{\phi\phi}$,  probes the matter \emph{perturbations} at redshift $z \sim 2$. It is very sensitive to the growth of structure and hence one of the tools for discerning between the $\Lambda$CDM model and its alternatives \cite{Perotto:2006rj,Calabrese:2009tt}, such as the interacting dark-sector models \cite{DeBernardis:2011iw}.  Thus,  as a novelty and our second aim, we test the ability of the publicly available Planck 2013 $C_\ell^{\phi\phi}$ likelihood to further constrain the interacting dark sector.\footnote{%
Although not explicitly clear, the Planck 2015 lensing likelihood (which has not been made publicly available as of 16th March 2015) may have been used in a different interacting dark-sector model in a Planck 2015 paper \cite{Ade:2015rim}, where a coupled dark energy model was constructed following \cite{Pettorino:2013oxa}.}

The Planck temperature CMB data (without lensing likelihood) have already been used also in constraining the interacting models with scalar field dark energy \cite{Pettorino:2013oxa}, type (1) interactions $Q_{\rm c} = H\xi\rhode$ \cite{Salvatelli:2013wra,Bolotin:2013jpa, Yang:2014gza, yang:2014vza, Li:2014cee, Yang:2014hea},  $Q_{\rm c} = H  (\xi_1\rhoc + \xi_2 \rhode)$  \cite{Costa:2013sva, Abdalla:2014cla, Pu:2014goa}, $Q_{\rm c} = H\xi\rhode\rhoc/(\rhode+\rhoc)$ \cite{Li:2013bya}, and type (2) interactions in \cite{Zhang:2013lea} --- but \cite{Zhang:2013lea} neglects the perturbations and derives only background constraints, so it does not significantly overlap with our work. Yet a few other couplings are introduced in \cite{Xu:2011qv,Baldi:2012kt,Chimento:2013rya,G:2014mea,Wang:2014xca}, with moderate Bayesian evidence for an interacting vacuum model \cite{Salvatelli:2014zta}, and steps toward more generic models are presented in \cite{Pourtsidou:2013nha,Boehmer:2015kta,Boehmer:2015sha,Gleyzes:2015pma,Skordis:2015yra}.

\section{Background evolution}

As we will study interaction rates that are smaller than today's expansion rate, $|\Gamma|/H_0 < 1$, the effect on the evolution of $\rhoc$ is mild and until very recently $\rhoc \propto a^{-3}$ to high accuracy. In contrary, the continuity equations \eqref{eq:cont_rhoc} and \eqref{eq:cont_rhode} with our chosen interaction term \eqref{eq:ourQc} may lead to a negative dark-energy density, $\rhode$, at some point of the evolution of the Universe, see e.g. \cite{Valiviita:2008iv,Valiviita:2009nu}. As the nature and origin of dark energy is still a mystery, this behaviour could be acceptable. However, at the moment of zero crossing from positive values to the negative ones, the perturbation equations are singular \cite{Majerotto:2009np}. Thus we will study only those models where $\rhode$ is positive today and in the past. (It could cross zero in the future, but this is not a problem, since in the lack of observations from the future we do not know whether perturbations might start growing rapidly.) Here we briefly discuss the background evolution.

We define a ratio of the energy densities,
\begin{equation}
r = \frac{\rhode}{\rhoc}\,,
\end{equation}
and employ equations \eqref{eq:cont_rhoc}, \eqref{eq:cont_rhode} and \eqref{eq:ourQc} to find a differential equation
\begin{equation}
r' = \left( \frac{a}{\aref}\Gamma - 3\h\wde \right)r + \frac{a}{\aref}\Gamma\,.
\label{eq:rprime}
\end{equation}
Had we studied a type (1) model with $Q_{\rm c} = -\Gamma H \rho_{\mr{c}}$, the resulting equation, $r' = \h(\Gamma-3\wde)r + \h\Gamma$, would have been trivial to solve analytically. This mathematical simplicity seems to have been one of the reasons (although a questionable one) for the popularity of type (1) models in the literature. In our type (2) model, with  $Q_c = -\Gamma \rho_{\mr{c}}$, the missing $\h$ from the $\Gamma$ terms in \eqref{eq:rprime} makes full analytic solution impossible, and we need to solve the evolution of $\rhoc$, $\rhode$, and $\h$ numerically. However, assuming radiation or matter domination, i.e., a known form for $\h$, we can find analytic solutions.

In the early universe, during radiation domination, $\h = \tau^{-1}$ and $a/\aref = \tau/\tauref$. With the initial condition $r(\tauref) = \rref$, we find
\begin{eqnarray}
r(\tau) & = & e^{(\Gamma/\tauref)(\tau^2 - \tauref^2)/2} \left( \frac{\tau}{\tauref}\right)^{-3\wde}\rref \ + \  e^{(\Gamma/\tauref)(\tau^2/2)} \left(\frac{\tau^2}{2}\frac{\Gamma}{\tauref}\right)^{-3\wde/2} \times 
\nonumber\\
&& \left[ G \left( \textstyle\frac{3}{2}\wde+1,\frac{\Gamma}{\tauref}\frac{\tauref^2}{2}\right) - 
     G \left( \textstyle\frac{3}{2}\wde+1,\frac{\Gamma}{\tauref}\frac{\tau^2}{2}\right)\right]\,,
\label{eq:rtau}
\end{eqnarray}
where $G$ is the incomplete Gamma function, $G(a,b) = \int_b^\infty x^{a-1} e^{-x} dx$. We choose $\tauref$ to be a later time than $\tau$, $\tauref > \tau$, and we assume that $\rhoc(\tauref)$ and $\rref$ are positive. Now we figure out whether $r$, and thus $\rhode$, is positive at all $\tau < \tauref$. The first term in \eqref{eq:rtau} is clearly positive. The second term, which involves the difference of incomplete Gamma functions, can be written as
\begin{equation}
  e^{(\Gamma/\tauref)(\tau^2/2)}   \int_{\frac{\Gamma}{\tauref}\frac{\tauref^2}{2}}^{\frac{\Gamma}{\tauref}\frac{\tau^2}{2}} \left(\frac{\tau^2}{2}\frac{\Gamma}{\tauref x}\right)^{-3\wde/2} e^x dx\,.
\end{equation}
The term in front of the integral is always positive. If $\Gamma < 0$, the integral is from a more negative number to a less negative number, while the integrand itself is positive (and real) in the whole range. Therefore, with $\Gamma < 0$ we have $r(\tau) > 0$ (at least) for $0 < \tau < \tauref$. If $\Gamma > 0$, then the integral is from a larger positive number to a smaller positive number, while the integrand itself is again positive. Since the integration is to a ``wrong'' direction, the result is negative. It is more negative the larger the $\Gamma$ is and the further back in time we go, and in the limit $\tau \rightarrow 0$ the integral diverges, if $-3\wde/2 \ge 1$. Thus with $\Gamma > 0$ and $\wde \le -2/3$, the integral approaches $-\infty$, when $\tau \rightarrow 0$. This means that for $\Gamma > 0$ we can find a positive solution for $r$, in the whole range $0 < \tau < \tauref$ only if $\wde > -2/3$ and $|\Gamma|$ is small enough.

During matter domination $\h = 2\tau^{-1}$ and $a/\aref = (\tau/\tauref)^2$, where $\tauref$ is now some later time at matter domination than $\tau$. Now we find
\begin{eqnarray}
r(\tau) & = & e^{(\Gamma/\tauref^2)(\tau^3 - \tauref^3)/3} \left( \frac{\tau}{\tauref}\right)^{-6\wde}\rref \ + \  e^{(\Gamma/\tauref^2)(\tau^3/3)} \left(\frac{\tau^3}{3}\frac{\Gamma}{\tauref^2}\right)^{-2\wde} \times 
\nonumber\\
&& \left[ G \left( \textstyle2\wde+1,\frac{\Gamma}{\tauref^2}\frac{\tauref^3}{3}\right) - 
     G \left( \textstyle2\wde+1,\frac{\Gamma}{\tauref^2}\frac{\tau^3}{3}\right)\right]\,.
\label{eq:rtauMD}
\end{eqnarray}
Similar arguments as in the radiation dominated case show that for $\Gamma < 0$, $r$ is positive. If $\Gamma > 0$ and $-2\wde \ge 1$, i.e., $\wde \le -1/2$, then the term, which involves the difference of incomplete Gamma functions, would approach $-\infty$ as $\tau \rightarrow 0$. However, we should remember that the matter dominated solution does not hold at the earliest times. So, if $\Gamma$ is suitably close to zero, $r$ may stay positive even if $-2/3 < \wde \le -1/2$.

\begin{table}[t]
\begin{tabular}{l|lll}
\hline
\hline
                                        &  $\wde < -1$                             &  $-1 < \wde\le-2/3$             &   $-2/3 < \wde < 1/3$  \\
\hline
 $\Gamma > 0$               &   negative at early times             &  negative at early times          & positive, if $|\Gamma|$ small enough   \\
 $\Gamma < 0$               & positive                                      & positive                                  & positive     \\
\hline
\hline
\end{tabular}
\caption{The sign of $\rhode$ between $\tau = 0$ and today, when we assume that today $\rhode$, $\rhoc$, and $\h$ are positive. Only models with positive $\rhode$ are studied in this paper. Here $\wde$ refers to the dark energy equation of state parameter at radiation and (early) matter dominated eras. \label{tab:bg}}
\end{table}

In table \ref{tab:bg} we summarize the background evolution. This behaviour is confirmed by our full numerical solutions. In the following, we parametrize the variable equation of state parameter of dark energy as $\wde = w_0 - w_a(1 - a)$ \cite{Chevallier:2000qy, Linder:2002et}, and rewrite it 
\begin{equation}
\wde = w_0 a + w_e (1 - a),
\label{eq:varyingwde}
\end{equation}
where $w_e = w_0 + w_a$ denotes the early time value of $\wde$ and $w_0$ the late time value. 

\begin{figure}
\centering
\includegraphics{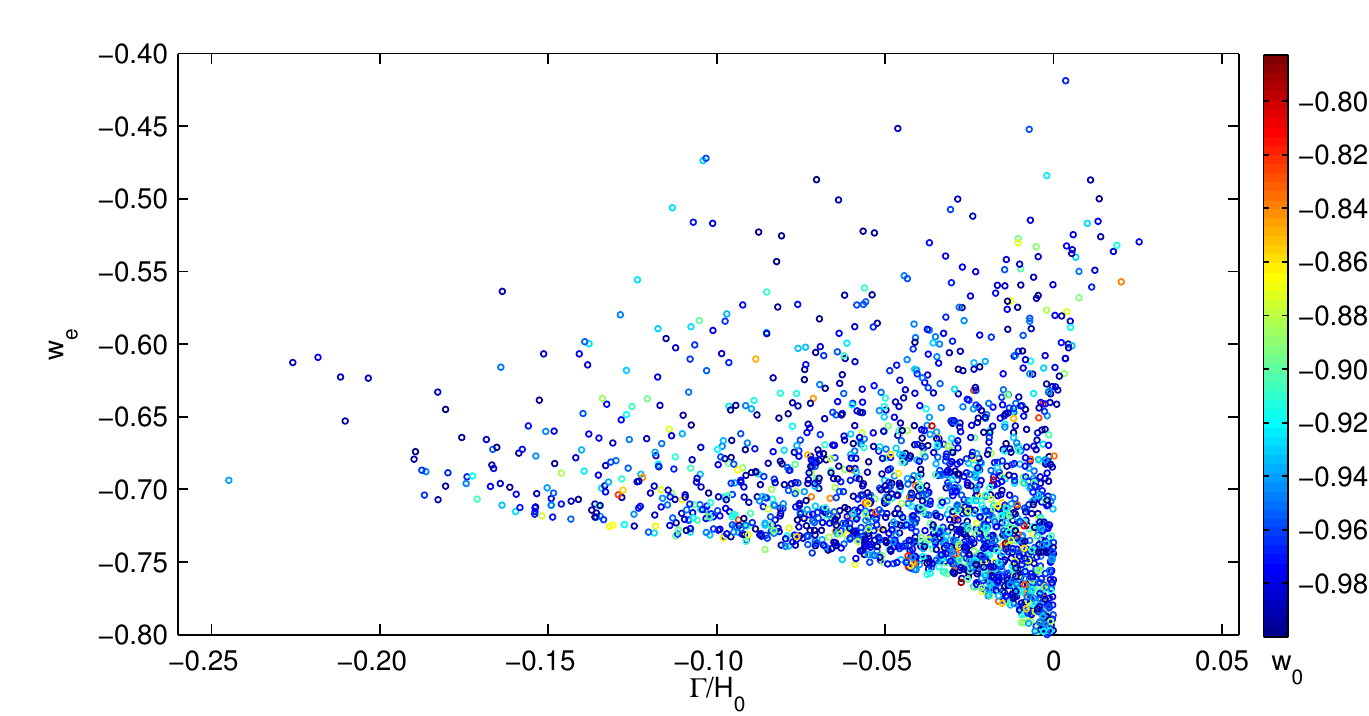}
\caption{{\bf The $w_e$ volume effect  in the non-phantom models ($\wde > -1$).} The points indicate samples from our Monte Carlo Markov Chains with the data combination CMB+BAO. The colour scale shows the value of $w_0$ for each sample. Whilst marginalizing (integrating) over the $w_e$ direction, the models with a positive $\Gamma$ (energy transfer from CDM to dark energy) receive much less weight than the ones with a negative $\Gamma$ (energy transfer from dark energy to CDM).\label{w0_we_Gamma_CMBandBAO_3D}}
\end{figure}

When analysing the results, one should bear in mind that, in the phantom case, only negative values of $\Gamma$ are possible; see the first column of table \ref{tab:bg}. Although both negative and positive values of $\Gamma$ are possible in the non-phantom case, the theoretical prior set by demanding positive definiteness of the past background-energy densities leads to much larger parameter-space volume of models with a negative $\Gamma$. As the last column of table \ref{tab:bg} indicates, in the non-phantom models with positive $\Gamma$ we must have $\wde > -2/3$ at early times and simultaneously $|\Gamma|$ small enough, in order to have the required background behaviour. Moreover, the perturbations "blow-up" if $w_e < -0.8$ in the non-phantom models \cite{Majerotto:2009np}. This adds another parameter≈ß-space volume effect to the analysis. The combination of these two effects is demonstrated in figure \ref{w0_we_Gamma_CMBandBAO_3D}.

\section{Perturbation evolution}

Since the model is phenomenological, the perturbation equations do not follow directly from variation of a given action. We need to do some choices and keep in mind that these can not be completely arbitrary as the formulation should be covariant. A good example is type (1) models where $H$ should not be treated as an average expansion rate, but instead $\delta H$ should be taken into account. For interaction \eqref{eq:ourQc} plausible perturbation equations were first derived in \cite{Valiviita:2008iv}. We will treat the perturbations in the same way as in \cite{Valiviita:2008iv} and \cite{Majerotto:2009np}, and assume a constant sound speed, $c_s^2 = 1$, for the dark energy. Here we only repeat the end result for the evolution of the dark energy and dark matter density contrast and their velocity potentials,
 \begin{eqnarray}
&& \delta'_{\mr{de}} + 3\mathcal
H(1-w_{\mr{de}})\delta_{\mr{de}} + (1+w_{\mr{de}})\left[\theta_{\mr{de}} +k^2
(B-E')\right] + 9\mathcal H^2(1-w_{\mr{de}}^2)\frac{\theta_{\mr{de}}}{k^2}
+3 \h w'_{\mr{de}} \frac{\theta_{\mr{de}}}{k^2}\nonumber\\
& & \quad -3(1+w_{\mr{de}})\psi' = 
a\Gamma \frac{\rhoc}{\rhode}\left[\delta_{\mr{c}}
-\delta_{\mr{de}} + 3\mathcal
H (1-w_{\mr{de}})\frac{\theta_{\mr{de}}}{k^2}+\phi\right]\!,\label{eq.delta'de_ourQ} \\
&& \theta'_{\mr{de}} -2 \mathcal H\theta_{\mr{de}}
-\frac{k^2}{(1+w_{\mr{de}})}\delta_{\mr{de}} - k^2\phi =
\frac{a\Gamma}{(1+w_{\mr{de}})}\frac{\rhoc}{\rhode}
\left(\theta_{\mr{c}}-2\theta_{\mr{de}}
\right)\,, \label{eq.theta'de_ourQ}\\
\label{eq.delta'c_ourQ} && \delta'_{\mr{c}} +\theta_{\mr{c}} + k^2 (B-E')
-3\psi' =-
a\Gamma\phi\,, \\
&& \theta'_{\mr{c}} + \mathcal H \theta_{\mr{c}} -k^2\phi = 0\,,
\label{eq.theta'c_ourQ}
 \end{eqnarray}
where $\phi$, $\psi$, $E$, and $B$ are scalar metric perturbations of spatially flat Friedmann-Robertson-Walker metric;
\begin{equation}
\label{eq.gengauge} 
ds^2 = a^2\Big\{ -(1+2\phi)
d\tau^2+2\partial_iB\,d\tau dx^i + \Big[(1-2\psi)\delta_{ij}+
2\partial_i\partial_j E\Big]dx^idx^j\Big\}\,. 
\end{equation}
Along with the modified background evolution equations, we have implemented the perturbation equations \eqref{eq.delta'de_ourQ}--\eqref{eq.theta'c_ourQ}, into Code for Anisotropies in the Microwave Background (\texttt{CAMB}) in synchronous gauge where $B = E = 0$, and the velocity perturbations take the form $\theta = -k^2v$. 
In the last term on the first line of equation \eqref{eq.delta'de_ourQ} we have, according to \eqref{eq:varyingwde},   $w'_{\mr{de}} = (w_0 - w_e) a' = -w_a a' = -w_a a \mathcal H$.

\section{Data and methods}

To determine the posterior probability density function (pdf) of the parameters of our interacting dark-sector model, as well as the non-interacting model for a reference, we perform a full Monte Carlo Markov Chain (MCMC) likelihood analysis, using our modified versions of \texttt{CosmoMC} \cite{Lewis:2002ah,Lewis:2013hha} and the \texttt{CAMB} \cite{Lewis:1999bs} Boltzmann code. The data sets used in the analysis include the Planck 2013 temperature anisotropy \cite{Ade:2013kta} and lensing data \cite{Ade:2013tyw}, and three baryon acoustic oscillation (BAO) measurements \cite{Beutler:2011hx,Padmanabhan:2012hf,Anderson:2012sa}, as described on page \pageref{page:BAO}. In all the analysis we supplement Planck temperature data with  the nine-year Wilkinson Microwave Anisotropy Probe (WMAP) polarization data \cite{Bennett:2012zja}, as done in \cite{Ade:2013zuv}, and label this combination as ``CMB''. For finding the best-fitting parameter values and the corresponding $\chi^2$ we use Bound Optimization BY Quadratic Approximation (\texttt{BOBYQA}) as provided in the \texttt{cosmomc} package, but with increased accuracy settings.

In the MCMC scans and in the best-fit searches we assume spatially flat geometry of the Universe and vary the following cosmological parameters: the physical baryon and CDM densities today ($\omega_b = h^2\Omega_b$ and $\omega_c = h^2\Omega_c$), the Hubble parameter, i.e., the expansion rate today ($H_0$ in km$\,$s$^{-1}\,$Mpc$^{-1}$), the optical depth due to reionization ($\tau$), the dark energy equation of state parameter today and at early times ($w_0$ and $w_e$), and the spectral index and logarithm of the amplitude of primordial scalar perturbations ($n_s$ and $\ln{A_s}$). From these we can calculate various derived parameters, such as the dark energy density parameter today ($\Omega_{\mathrm{de}}$) and the current age of the Universe (Age). In addition we vary the nuisance parameters of the Planck likelihood code as per \cite{Ade:2013zuv}, and in the interacting dark-sector case also the interaction rate in units of today's expansion rate, $\Gamma/H_0$, in the range (-1,1).

\section{Results}

We start with the non-phantom interacting model, which turns out not to improve the fit to the data over the non-interacting model. Then we proceed to the more interesting case of phantom models where the interaction can moderately improve the fit to the data, and the interaction can significantly affect the inferred values of standard cosmological parameters.

\subsection{Non-phantom model ($\wde > -1$)}

\begin{figure}
\includegraphics[width=\textwidth]{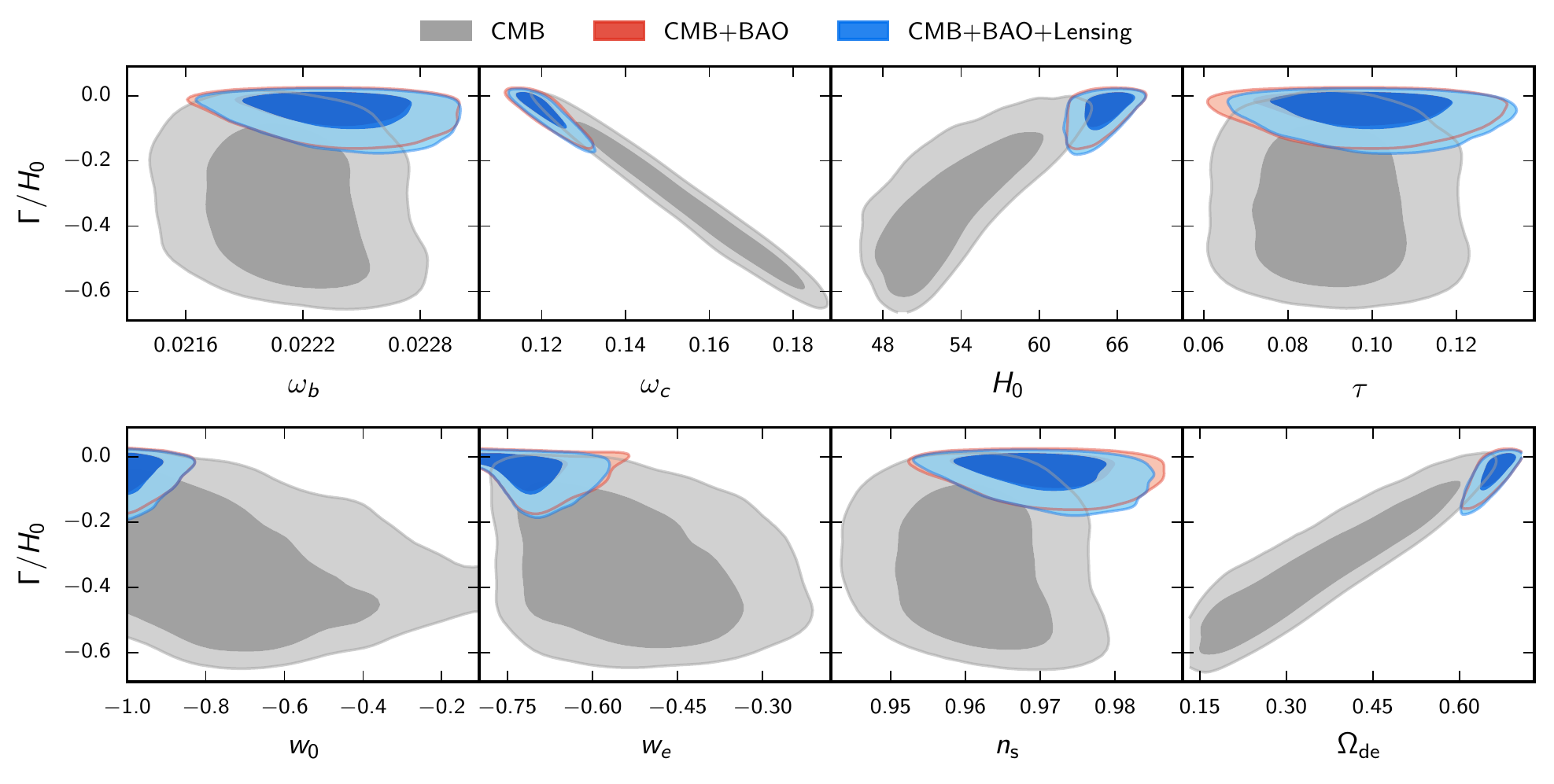}
\caption{{\bf The non-phantom interacting model ($\wde  > -1$).}  68\% and 95\% CL regions
with CMB (gray), CMB+BAO (red), and CMB+BAO+lensing (blue) data.\label{FigNoPhantom_2D}}
\end{figure}
\begin{figure}
\includegraphics[width=\textwidth]{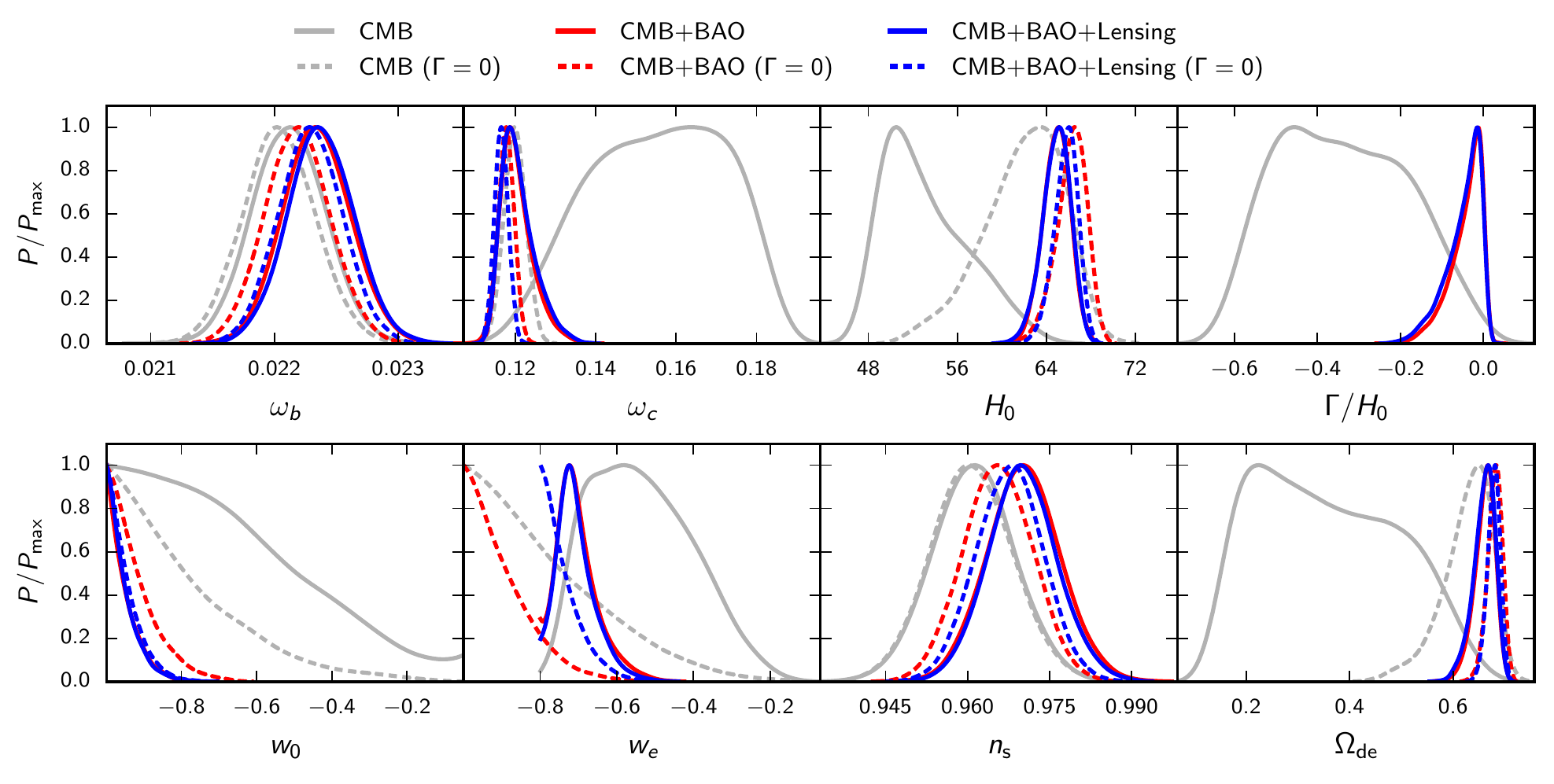}
\caption{{\bf Comparison of the non-phantom models ($\wde > -1$).} 1d marginalized posterior probability densities for the interacting dark-sector model (solid lines) and for the non-interacting ``standard'' $w$CDM model (dashed lines). In the non-interacting model, with the CMB and CMB+BAO data, we allow the early-time dark energy equation of state parameter, $w_e$, to vary between -1 and +1/3, whereas with the CMB+BAO+lensing data we allow only the same range, $w_e\in(-0.8,+1/3)$, which is possible in the interacting model (see figure \ref{w0_we_Gamma_CMBandBAO_3D}), in order to provide a different perspective for comparing the models.\label{FigNoPhantom_1D}}
\end{figure}

In figure \ref{FigNoPhantom_2D} we represent 68\% and 95\% confidence level (CL) contours in 2d slices of various standard cosmological parameters versus the interaction rate. The current CMB data alone leave almost as long degeneracy line in ($\omega_c$,$\,\Gamma/H_0$) plane as already found in \cite{Valiviita:2009nu} with the WMAP five-year \cite{Dunkley:2008ie} and the Arcminute Cosmology Bolometer Array Receiver (ACBAR) \cite{Reichardt:2008ay} data. This is natural, since CMB probes $\omega_{c\ast}$, i.e., the physical CDM density parameter at last scattering, not the density today. A positive $\Gamma$ implies that the CDM density decreases faster than in the non-interacting case. So, in order to fit well to the acoustic peak structure of the CMB data, i.e., to have the same $\omega_{c\ast}$ in the interacting model (with a positive $\Gamma$) as in the non-interacting model, we need a smaller $\omega_c$ today. On the other hand, with a negative $\Gamma$ we need a larger $\omega_c$ today, in order to have a correct  $\omega_{c\ast}$.

However, we immediately notice also some differences to \cite{Valiviita:2009nu}. Most notably, the Planck data disfavour positive values of $\Gamma$ over the negative ones more strongly than the WMAP data (see, e.g. the second panel of figure \ref{FigNoPhantom_2D}, the fourth panel of figure \ref{FigNoPhantom_1D}, and compare also figure \ref{w0_we_Gamma_CMBandBAO_3D} to a similar one in the appendix of \cite{Valiviita:2009nu}). A reason for this is that a negative $\Gamma$ leads to a larger $\omega_c$, and hence to a smaller (derived parameter) $\Omega_{\mathrm{de}}$. This smaller $\Omega_{\mathrm{de}}$ in turn causes much reduced integrated Sachs-Wolfe (ISW) effect compared to the non-interacting model or to the model with a positive $\Gamma$. At the lowest multipoles ($\ell \sim 2$--$40$), the Planck data have less temperature angular power (compared to the high multipoles) than the non-interacting $\Lambda$CDM model predicts. As the ISW effect adds power to the lowest multipoles the interacting model with a negative $\Gamma$ can significantly improve the fit in this range, whereas a positive $\Gamma$ leads to a worse fit.

As already mentioned, some non-CMB data are needed to break the ($\omega_c$,$\,\Gamma$) degeneracy, and a powerful choice is the BAO data; see, e.g., figure 4 in \cite{Valiviita:2009nu}. The BAO data would lead to almost orthogonal constraints compared to the gray CMB alone curves in the top-left corner in the second panel of figure \ref{FigNoPhantom_2D}, forcing the combined CMB+BAO good-fit region near to the zero interaction. Adding the (Planck 2013) CMB lensing data does not modify the constraints. 

In figure \ref{FigNoPhantom_1D} we show 1d marginalized pdfs for all the primary parameters (except $\tau$ and $\ln A_s$) and for one derived parameter $\Omega_{\mathrm{de}}$, comparing the results obtained in the interacting model to the ones in the non-interacting model. This comparison is not straightforward, since the CMB data (and in particular CMB+BAO or CMB+BAO+lensing) would favour $w_e \sim -1$, which is excluded in the interacting model (recall figure \ref{w0_we_Gamma_CMBandBAO_3D} which depicts the theoretical prior on $w_e$). The consequence is that in the non-interacting model the most favoured values of $w_e$ fall to the range $-1 < w_e < -0.8$, which is forbidden in the interacting model. In order to demonstrate more directly what kind of effects the interaction has on the determination of the other parameters, we have made the MCMC run with CMB+BAO+lensing data in the non-interacting model with a prior $-0.8 \le w_e < 1/3$. Hence, the solid blue ($\Gamma \neq 0$) and dashed blue ($\Gamma = 0$) lines indicate best the differences between the interacting and non-interacting model: in the interacting model a smaller $H_0$ and $\Omega_{\mathrm{de}}$ and larger $\omega_c$ are allowed (actually by any combination of data studied here). Table \ref{bestfits_NoPhantom} summarizes the best-fitting values and 68\% CL intervals of parameters, as well as, the shifts of the mean values of parameters compared to the non-interacting model in terms of the standard deviation ($\sigma$) of each parameter in the non-interacting model. For each parameter the first line represents the results in the interacting model and the second line in the non-interacting model. As was obvious from the figures, with CMB alone $\omega_c$ shifts upward by more than $13\sigma$ and this is reflected in a shift downward by $H_0$ ($-2.4\sigma$) and $\Omega_{\mathrm{de}}$ ($-5.5\sigma$). Adding the BAO or BAO+lensing data, restricts these shifts to $-1.6\sigma$. The least sensitive parameters are $\tau$ and $\ln A_s$; these shift by less than $0.3\sigma$.

\begin{table}
\centering
\begin{tabular}{lcccccc}
\hline
\hline
   & CMB  & CMB+BAO & \multicolumn{2}{c}{CMB+BAO+lensing} \\
\hline
WMAP polarization (\texttt{lowlike}) & 2.16 & 1.24 & 0.00 & {\bf -0.03}  \\
Planck low-$\ell$ TT, $2\le\ell\le 49$ (\texttt{commander}) & -1.24 & -1.32 & -0.10 & {\bf -0.16}  \\
Planck high-$\ell$ TT, $50\le\ell\le 2500$  (\texttt{CAMspec}) & -2.02 & 0.67 & 1.51 & {\bf 0.05}  \\
Three BAO measurements \cite{Beutler:2011hx,Padmanabhan:2012hf,Anderson:2012sa} & -- & 1.06 & 0.49 & {\bf 0.04}  \\
Planck CMB lensing  ($C_\ell^{\phi\phi}$) & -- & -- & -0.49 & {\bf 0.00}  \\
\hline
$\Sigma$ & -1.10 & 1.64 & 1.41 & {\bf -0.11} \\
\hline
\hline
\end{tabular}
\caption{{\bf Difference of the the best-fit $\chi^2$ between the interacting and non-interacting non-phantom ($\wde > -1$) model.} The first three columns indicate the difference to the non-interacting model with $-1 < w_e < 1/3$, while the last column (highlighted by bold face) gives the difference to the non-interacting model with  $-0.8 \le w_e < 1/3$. Negative values mean that the interacting model provides a better fit to the indicated data combination than the non-interacting one.\label{chi2_NoPhantom}}
\end{table}

Finally, we check whether the non-phantom interacting dark-sector model improves the fit to the data; see table \ref{chi2_NoPhantom}. Once we break the  ($\omega_c$,$\,\Gamma)$ degeneracy with BAO or BAO+lensing, the interacting model actually worsens the fit to the data compared to the non-interacting model, if we allow in the non-interacting model $-1 < w_e < 1/3$. This is due to the fact that the combinations CMB+BAO or CMB+BAO+lensing strongly favour $w_e < -0.8$, as already mentioned. In the last column we allow for the non-interacting model only values $w_e \ge -0.8$, and in this case we naturally find that the interacting model with its one extra parameter compared to the non-interacting model indeed improves the fit slightly, $\Delta\chi^2 = -0.11$. The largest improvement, $\Delta\chi^2 = -0.16$, comes from the Planck low-$\ell$ TT data due to the decreased $\Omega_{\mathrm{de}}$ and thus a reduced ISW contribution. The overall conclusion is that the data do not favour the non-phantom interacting model.

\subsection{Phantom model ($\wde < -1$)}

In the phantom models the interacting and non-interacting cases are directly comparable, since all the standard parameters can have the same prior in both cases. In particular, we choose the uniform priors $-3 < w_0 < -1$ and $-3 < w_e < -1$. Now the comparison of $\chi^2$ of the best-fitting interacting and non-interacting models shows a clear preference for the interacting model; see table \ref{chi2_Phantom}. With CMB+BAO+lensing data the largest improvement comes again from the low-$\ell$ Planck TT data, where the ISW effect can be reduced in the interacting model.

\begin{table}
\centering
\begin{tabular}{lccccc}
\hline
\hline
                        & CMB  & CMB+BAO & CMB+BAO+lensing \\
\hline
WMAP polarization (\texttt{lowlike})  & 1.48 & 0.88 & 0.95 \\
Planck low-$\ell$ TT, $2\le\ell\le 49$ (\texttt{commander})  & -1.14 & -2.25 & -2.10 \\
Planck high-$\ell$ TT, $50\le\ell\le 2500$  (\texttt{CAMspec}) & -1.05 & -2.37 & -0.41 \\
Three BAO measurements \cite{Beutler:2011hx,Padmanabhan:2012hf,Anderson:2012sa}  & -- & -0.14 & -0.13 \\
Planck CMB lensing ($C_\ell^{\phi\phi}$)  & -- & -- & -0.47 \\
\hline
$\Sigma$ & -0.71 & -3.87 & -2.17 \\
\hline
\hline
\end{tabular}
\caption{{\bf Difference of the the best-fit $\chi^2$ between the interacting and non-interacting phantom ($\wde < -1$) model.}\label{chi2_Phantom}}
\end{table} 

\begin{figure}
\includegraphics[width=\textwidth]{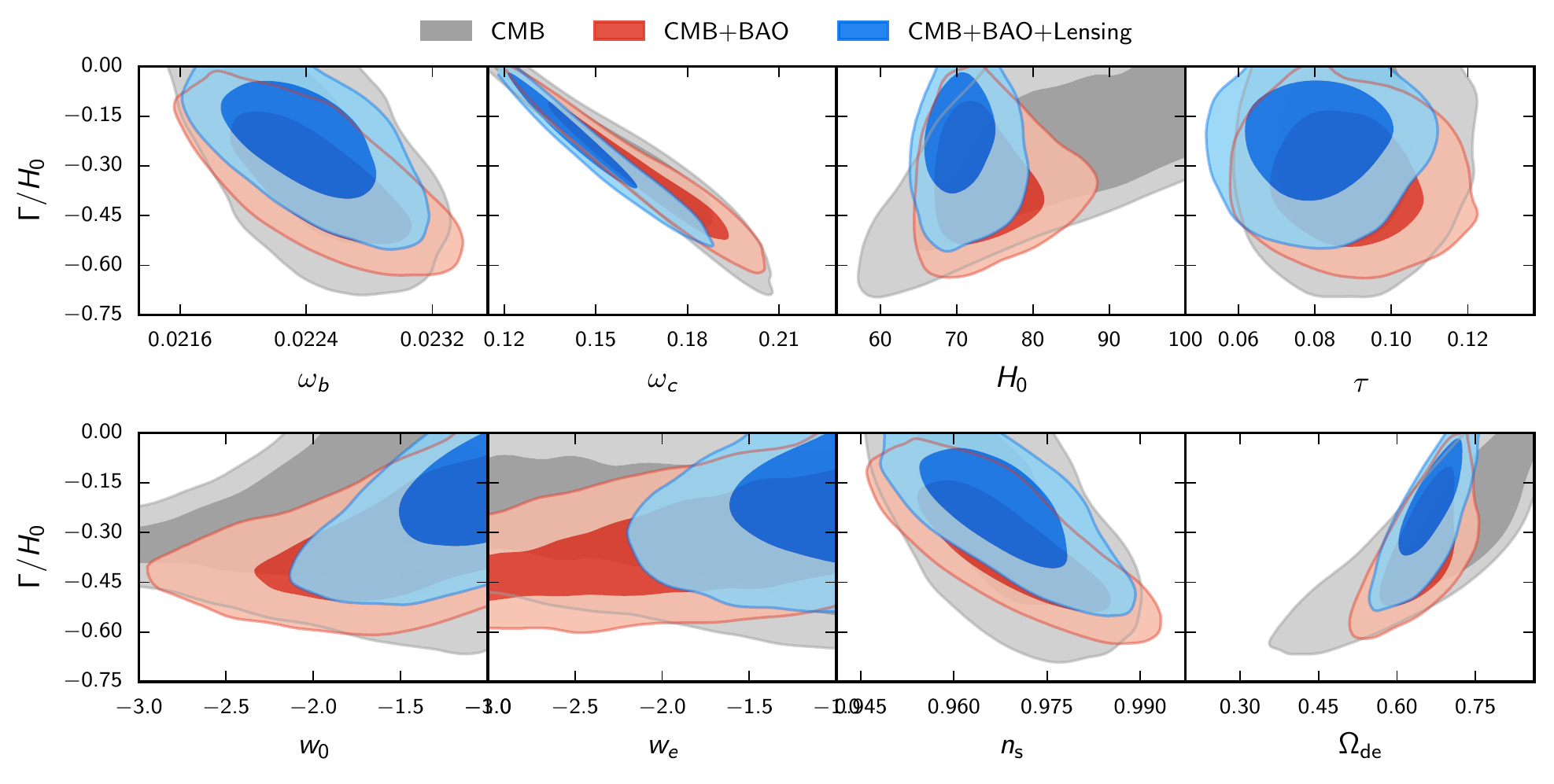}
\caption{{\bf The phantom interacting model ($\wde < -1$) with CMB, BAO, and lensing data.}\label{FigPhantom_2D}}
\end{figure}

\begin{figure}
\includegraphics[width=\textwidth]{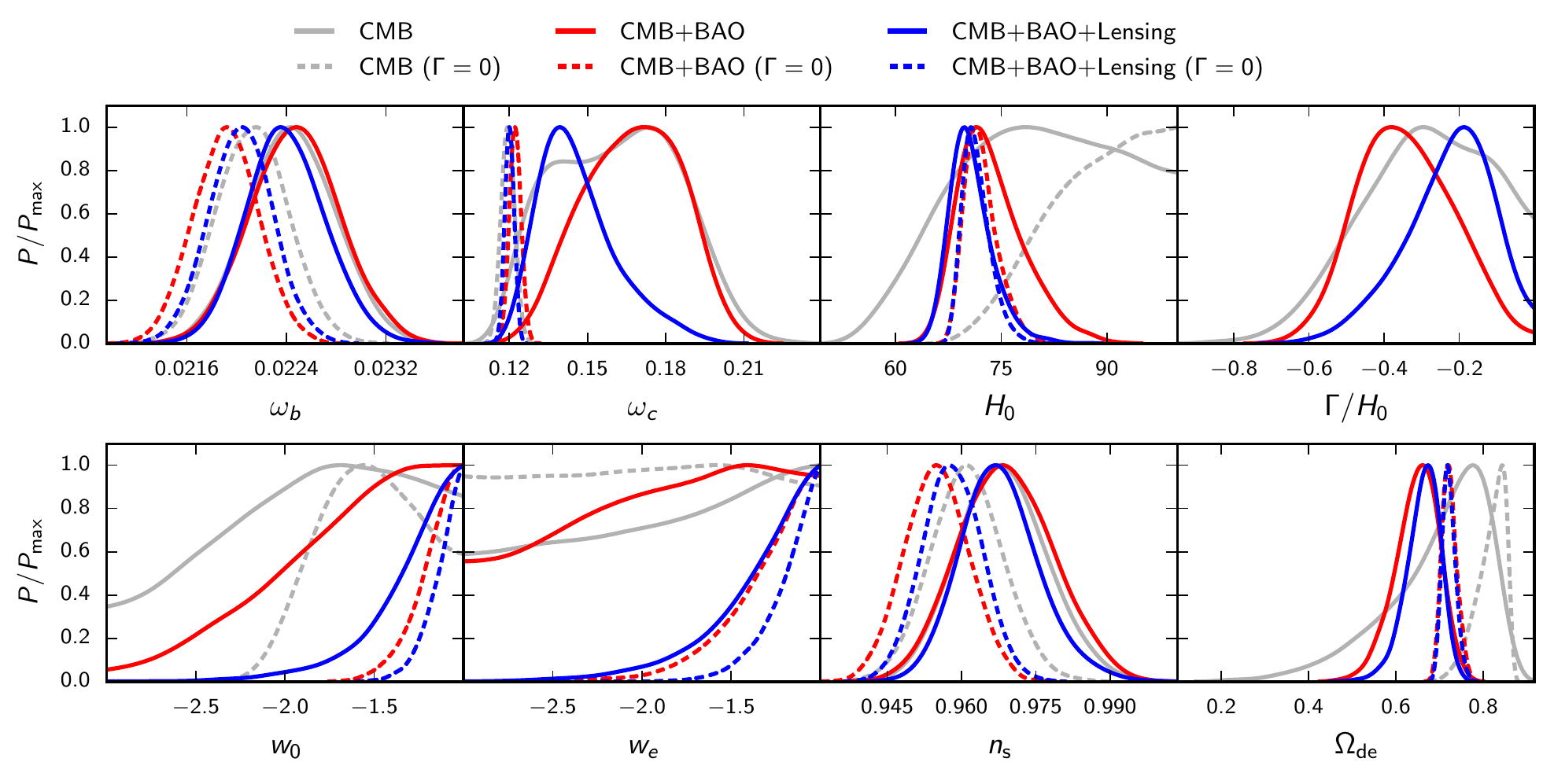}
\caption{{\bf The phantom model ($\wde < -1$) with CMB, BAO, and lensing data.}\label{FigPhantom_1D}}
\end{figure}

In the phantom model the 2d slices reveal many interesting things; see figure \ref{FigPhantom_2D}. Firstly, with CMB+BAO data, the non-interacting model is outside the 95\% CL region, except in the ($\tau$,$\,\Gamma$) and  ($w_0$,$\,\Gamma$) figures, where $\Gamma=0$ is at the border of this region. The marginalized 1d pdf of $\Gamma/H_0$ in the fourth panel of figure \ref{FigPhantom_1D} thus shows a clear peak at as high an interaction rate as $\Gamma/H_0 = -0.4$, about $3\sigma$ away from zero. According to table \ref{bestfits_Phantom} even the mean, $\Gamma/H_0=-0.34$, is almost $3\sigma$ away from zero. Therefore, with CMB+BAO data we might conclude that they moderately favour the interacting phantom model, giving an improvement of $3.9$ for the best fit $\chi^2$ with only one extra parameter.

Adding the CMB lensing data to the constraint budget, restores the non-interacting model to or near to the border of the 68\% CL region; see the blue curves in figure \ref{FigPhantom_2D}. We underline that
 there are two ``separate'' effects in the game: (a) An indirect improvement of the constraints for $\Gamma$ due to the improved constraints of the equation of state parameters $w_0$ and $w_e$, which is visible also in the non-interacting model --- compare the sequence of the gray, red, and blue dashed curves in  figure \ref{FigPhantom_1D} ($w_0$ and $w_e$ panels). In the interacting model, adding BAO (red solid) to CMB alone (gray solid) does not change the pdf of $w_e$ almost at all, but the addition of the lensing data (blue solid) leads to a solid constraint on $w_e$ even. Moreover, figure \ref{FigPhantom_1D} and table \ref{bestfits_Phantom} reveal that (both in the non-interacting and interacting models) adding the lensing data shifts $\omega_c$ to smaller values than favoured by CMB+BAO. Recalling the ($\omega_c$,$\,\Gamma$) degeneracy, this causes a shift to a less negative interaction rate. (b) As explained in Introduction, the lensing data constrain the growth rate of perturbations (mainly around redshift $z \sim 2$). Since the interaction directly affects the evolution of perturbations, the lensing data hence probe also directly the interaction rate. The combined effect of (a) and (b) is to drive $\Gamma$ closer to zero. However, the 1d marginalized pdf of the interaction rate still peaks at a clearly non-zero value, $\Gamma/H_0 = -0.19$ (figure \ref{FigPhantom_1D}), the best-fitting value being $-0.12$ and the mean being $-0.23$, which is about $1.6\sigma$ away from zero. Nevertheless, this demonstrates how powerful the lensing data are in discerning between the standard $\Lambda$CDM (or in our case $w$CDM) and the alternative dark-energy or dark-sector models.


Unlike in the non-phantom model, in the phantom model allowing for an interaction affects dramatically the preferred values of some standard cosmological parameters even with CMB+BAO (or CMB+BAO+lensing) data. Naturally, the most affected is $\omega_c$, which shifts $20\sigma$ ($15\sigma$) upward. Table \ref{bestfits_Phantom} shows that this is reflected in $\Omega_{\mathrm{de}}$  by a $4\sigma$ shift downward; from the mean value of $\Omega_{\mathrm{de}} = 0.72$ to $0.65$ ($0.66$). The values of other parameters stay largely unaffected, except $n_s$ which moves $2\sigma$ upward. Despite of this, the scale-invariant primordial spectrum remains $3.3\sigma$ ($4.5\sigma$) disfavoured by the data in the interacting model, while it is $7\sigma$ disfavoured in the non-interacting $w$CDM model.

\section{Discussion}

We presented a numerical analysis of a phenomenological interacting dark-sector model, where dark energy was treated as a fluid with a time-dependent equation of state parameter, $\wde = w_0 a + w_e (1-a)$. We considered a model with energy transfer from dark matter to dark energy of the form $Q_c = -\Gamma\rho_{\mr{c}}$, and assumed the relative energy transfer rate $\Gamma$ to be a constant in time. We updated the previous analysis \cite{Valiviita:2009nu} of the non-phantom model $\wde > -1$, using the Planck 2013 data, and --- for completeness --- extended the analysis also to the phantom dark energy, $\wde < -1$. For the first time, in the analysis of the interacting dark sector, we utilized the CMB lensing potential reconstruction data ($C_\ell^{\phi\phi}$).

In the non-phantom models the non-interacting $w$CDM model ($\Gamma=0$) lied close to the peak of the posterior pdf; $-0.069 < \Gamma/H_0 < 0.005$ at 68\% CL and $-0.14 < \Gamma/H_0 < 0.02$ at 95\% CL, with CMB+BAO+lensing, and allowing for the interaction (and thus one extra parameter) did not improve the fit to the data. On the contrary, in the phantom models an interaction rate higher than 10\% of the expansion rate, $\Gamma/H_0 = -0.16$ ($-0.12$), led to an improvement of $\chi^2$ by $3.9$ ($2.2$) with CMB+BAO (CMB+BAO+lensing) data. The non-interacting model lied outside of the 95\% CL interval,  $-0.57 < \Gamma/H_0 < -0.10$ ($-0.46 < \Gamma/H_0 < -0.01$). Our results in the phantom model indicated that the lensing data have a great potential for discerning between interacting and non-interacting models, and considerably improve the constraints on $\Gamma$, as well as, on the dark energy equation of state parameter today and at early times, $w_0$ and $w_e$.

We focused on the effects that can be calculated within linear perturbation theory (by using our modified version of \texttt{CAMB}), and thus we used only CMB and BAO data, and in particular demonstrated the discerning power of the CMB lensing data.  Although the shape and amplitude of the matter power spectrum might provide some extra constraints, the use of it up to the smallest scales (largest wave numbers $k$) probed by the current large-scale structure observations would require detailed modeling of the non-linear effects that may differ from the non-interacting models, as shown in \cite{Baldi:2010vv,Li:2010re,Baldi:2010pq,Baldi:2011wa,Baldi:2011wy}. Yet there would remain ambiguity of the assumptions made on the bias function, i.e., how the dark matter traces the visible matter on the smallest observable scales. Going to even smaller scales would require, e.g.,  the use of Lyman-$\alpha$ forest data \cite{Baldi:2010ks}.

In the near future, the Dark Energy Survey (DES) \cite{Sevilla:2009zza} and later the Euclid \cite{Amendola:2012ys} data will offer an unprecedented discerning power.  Dark-sector interaction can increase the probability of very massive clusters occurring at high redshift \cite{Baldi:2010td} and existence of structures at very large scales \cite{Mota:2008nj}. Voids in the interacting models are emptier and contain less neutral hydrogen than in the $\Lambda$CDM model, and this effect might alleviate tensions between simulations and observations in voids \cite{Baldi:2010ks}. Further,  weak lensing has been shown to offer valuable constraints in the future \cite{Beynon:2011hw,DeBernardis:2011iw}. In this paper, we mentioned several times the modified ISW effect in the interacting dark-sector models. More direct ISW data (other than an increased low-$\ell$ temperature angular power which is impossible to distinguish from other effects) is provided by the cross-correlation of the CMB and large-scale structure maps; see, e.g, \cite{Giannantonio:2012aa,Ade:2013dsi,Soergel:2014sna,Ade:2015dva}. As already mentioned, an interaction also directly modifies the growth rate of matter perturbations \cite{Simpson:2010yt}, which can make it difficult to distinguish from modified gravity \cite{Wei:2008vw}. In the background level, the effects of interaction may be absorbed into a modified effective equation of state parameter $w_{\mathrm{de,eff}}$; see \cite{Valiviita:2008iv}. Hence an interacting model might seem like a non-interacting model with a different equation of state parameter. This degeneracy is studied in \cite{Avelino:2012tc}. Extensive summaries of the constraining or distinguishing power of various existing and future low-redshift (and) large-scale surveys are provided by 
\cite{Honorez:2010rr,Amendola:2011ie}.

In this paper we adopted a phenomenological ``top-down'' approach, i.e., introduced an energy(-momentum) transfer term to the background (and perturbation) continuity equations and studied whether such an extension of the $w$CDM model was favoured by the data. Another possibility would have been a ``bottom-up'' approach, i.e., identifying fundamental theories where an interaction arises and then studying a particular form of the interaction predicted by the theory. Recently, an interesting scenario, which leads naturally to an interacting dark sector, has been put forward \cite{Koivisto:2013fta}. The so called disformal coupling (see also \cite{Gleyzes:2015pma}) arises from Dirac-Born-Infeld actions in Type II string theories, when matter resides on a moving hidden sector D-brane. As a bonus, the coupling in this scenario can alleviate the coincidence problem of dark energy. Since the current data do not exclude interactions in the dark sector and indeed can accommodate quite large interaction rates, this theory offers interesting prospects for the future work.

Recently, progress in phenomenological or effective description has also been made by generalizing the parametrized post-Friedmann framework (PPF) \cite{Hu:2007pj} to the case of interacting dark energy \cite{Li:2014eha,Li:2014cee}. PPF can be used to describe very generic types of interaction \cite{Skordis:2015yra}, at the cost of up to 12 free functions or extra parameters. However, when applied to specific models the number of extra degrees of freedom reduces dramatically. For example, in the case of our model to just one, as shown in  \cite{Skordis:2015yra}.

\acknowledgments
This work was supported by the Academy of Finland grant 257989. 
We thank CSC --- IT Center for Science Ltd (Finland) for computational resources. 
Significant parts of the results have been achieved using the PRACE-3IP project (FP7 RI-312763) resource Sisu
at CSC.
The CMB and lensing likelihoods used in this work are described in \cite{Ade:2013kta,Ade:2013tyw} and available via Planck Legacy Archive. 
They are based on observations obtained with Planck, an ESA science mission with instruments and contributions directly funded by ESA Member States, NASA, and Canada.

\begin{landscape}
\begin{table}
\small
\begin{tabular} { l  l l l l l l }
\noalign{\vskip 3pt}\hline\noalign{\vskip 1.5pt}\hline\noalign{\vskip 5pt}
 \multicolumn{1}{c}{\bf } &  \multicolumn{2}{c}{\bf CMB} &  \multicolumn{2}{c}{\bf CMB+BAO} &  \multicolumn{2}{c}{\bf CMB+BAO+lensing} \\
\noalign{\vskip 3pt}\cline{2-7}\noalign{\vskip 3pt}

 Parameter &  Best fit &  68\% limits &  Best fit &  68\% limits &  Best fit &  68\% limits \\
\hline
{\boldmath$\omega_b       $} & $0.022169                  $ & $0.02212\pm 0.00029\quad(+0.3 \sigma)$ & $0.022200                  $ & $0.02235\pm 0.00027\quad(+0.6 \sigma)$ & $0.022298                  $ & $0.02238\pm 0.00026\quad(+0.3 \sigma)$ \\
\ \ \ ($\Gamma=0$) & $0.022096                  $ & $0.02204\pm 0.00029        $ & $0.022121                  $ & $0.02219\pm 0.00026        $ & $0.022275                  $ & $0.02229\pm 0.00025        $ \\
\hline

{\boldmath$\omega_c       $} & $0.1716                    $ & $0.155^{+0.021}_{-0.016}\quad(+13.2 \sigma)$ & $0.11788                   $ & $0.1204^{+0.0027}_{-0.0048}\quad(+1.4 \sigma)$ & $0.11735                   $ & $0.1208^{+0.0026}_{-0.0052}\quad(+2.5 \sigma)$ \\
\ \ \ ($\Gamma=0$) & $0.11971                   $ & $0.1199\pm 0.0027          $ & $0.11912                   $ & $0.1177\pm 0.0019          $ & $0.11709                   $ & $0.1165\pm 0.0017          $ \\
\hline

{\boldmath$H_0            $} & $51.80                     $ & $53.2^{+2.5}_{-5.0}\quad(-2.4 \sigma)$ & $66.73                     $ & $64.9^{+1.3}_{-1.1}\quad(-1.0 \sigma)$ & $67.16                     $ & $64.9^{+1.4}_{-1.1}\quad(-0.8 \sigma)$ \\
\ \ \ ($\Gamma=0$) & $67.06                     $ & $61.9^{+4.4}_{-2.8}        $ & $68.01                     $ & $66.2^{+1.6}_{-1.1}        $ & $67.20                     $ & $65.7^{+1.2}_{-0.90}       $ \\
\hline

{\boldmath$\tau           $} & $0.0903                    $ & $0.091^{+0.013}_{-0.014}\quad(+0.1 \sigma)$ & $0.0990                    $ & $0.097\pm 0.014\quad(+0.3 \sigma)$ & $0.0924                    $ & $0.099^{+0.012}_{-0.014}\quad(+0.2 \sigma)$ \\
\ \ \ ($\Gamma=0$) & $0.0853                    $ & $0.089^{+0.013}_{-0.014}   $ & $0.0892                    $ & $0.093^{+0.013}_{-0.014}   $ & $0.0932                    $ & $0.096^{+0.012}_{-0.014}   $ \\
\hline

{\boldmath$\Gamma/H_0     $} & $-0.475                    $ & $-0.34^{+0.15}_{-0.19}     $ & $-0.0007                   $ & $-0.043^{+0.048}_{-0.018}  $ & $-0.0024                   $ & $-0.049^{+0.054}_{-0.020}  $ \\
\ \ \ ($\Gamma=0$) & $0                         $ & $\mbox{---}                $ & $0                         $ & $\mbox{---}                $ & $0                         $ & $\mbox{---}                $ \\
\hline

{\boldmath$w_0            $} & $-0.835                    $ & $< -0.595\quad(+0.7 \sigma)$ & $-1.0000                   $ & $< -0.946\quad(-0.5 \sigma)$ & $-0.9990                   $ & $< -0.948\quad(-0.1 \sigma)$ \\
\ \ \ ($\Gamma=0$) & $-1.000                    $ & $< -0.758                  $ & $-1.0000                   $ & $< -0.911                  $ & $-0.9998                   $ & $< -0.941                  $ \\
\hline

{\boldmath$w_e            $} & $-0.419                    $ & $-0.532^{+0.099}_{-0.18}\quad(+1.6 \sigma)$ & $-0.7784                   $ & $-0.704^{+0.036}_{-0.063}\quad(+2.4 \sigma)$ & $-0.7987                   $ & $-0.706^{+0.033}_{-0.059}\quad(+0.7 \sigma)$ \\
\ \ \ ($\Gamma=0$) & $-0.901                    $ & $< -0.737                  $ & $-0.983                    $ & $< -0.880                  $ & $-0.7990                   $ & $< -0.728                  $ \\
\hline

{\boldmath$n_\mathrm{s}   $} & $0.9620                    $ & $0.9611\pm 0.0074\quad(+0.0 \sigma)$ & $0.9670                    $ & $0.9703\pm 0.0066\quad(+0.8 \sigma)$ & $0.9680                    $ & $0.9701\pm 0.0062\quad(+0.4 \sigma)$ \\
\ \ \ ($\Gamma=0$) & $0.9630                    $ & $0.9608\pm 0.0074          $ & $0.9634                    $ & $0.9656\pm 0.0062          $ & $0.9674                    $ & $0.9676\pm 0.0060          $ \\
\hline

{\boldmath$\ln(10^{10} A_\mathrm{s})$} & $3.0902                    $ & $3.090^{+0.024}_{-0.027}\quad(+0.1 \sigma)$ & $3.1045                    $ & $3.093\pm 0.027\quad(+0.1 \sigma)$ & $3.0897                    $ & $3.098\pm 0.025\quad(+0.2 \sigma)$ \\
\ \ \ ($\Gamma=0$) & $3.0821                    $ & $3.089\pm 0.025            $ & $3.0879                    $ & $3.090^{+0.025}_{-0.028}   $ & $3.0906                    $ & $3.094^{+0.023}_{-0.026}   $ \\
\hline

$\Omega_{\mathrm{de}}      $ & $0.278                     $ & $0.36^{+0.10}_{-0.19}\quad(-5.5 \sigma)$ & $0.6855                    $ & $0.661^{+0.024}_{-0.015}\quad(-1.4 \sigma)$ & $0.6904                    $ & $0.659^{+0.025}_{-0.016}\quad(-1.6 \sigma)$\\
\ \ \ ($\Gamma=0$) & $0.6846                    $ & $0.626^{+0.059}_{-0.030}   $ & $0.6946                    $ & $0.680^{+0.017}_{-0.012}   $ & $0.6914                    $ & $0.678^{+0.014}_{-0.011}   $\\
\hline

$\mathrm{Age}/\mathrm{Gyr} $ & $14.415                    $ & $14.33^{+0.21}_{-0.14}\quad(+3.3 \sigma)$ & $13.8243                   $ & $13.880\pm 0.049\quad(+1.4 \sigma)$ & $13.808                    $ & $13.882^{+0.047}_{-0.054}\quad(+0.8 \sigma)$\\
\ \ \ ($\Gamma=0$) & $13.810                    $ & $13.948^{+0.083}_{-0.14}   $ & $13.7785                   $ & $13.816^{+0.042}_{-0.049}  $ & $13.8093                   $ & $13.848^{+0.040}_{-0.045}  $\\
\hline
\hline
\end{tabular}
\caption{{\bf The non-phantom model ($\wde > -1$).} Best-fitting parameter values, mean and 68\% CL intervals for selected cosmological parameters (the primary MCMC parameters in bold, and derived parameters in non-bold face). For each parameter the first line is for the interacting dark-sector model and the second line for the non-interacting model ($\Gamma=0$). In parenthesis, the shift of the mean value compared to the non-interacting model is indicated in units of the standard deviation of the corresponding parameter of the non-interacting model.\label{bestfits_NoPhantom}}
\end{table}

\begin{table}
\small
\begin{tabular} { l  l l l l l l }
\noalign{\vskip 3pt}\hline\noalign{\vskip 1.5pt}\hline\noalign{\vskip 5pt}
 \multicolumn{1}{c}{\bf } &  \multicolumn{2}{c}{\bf CMB} &  \multicolumn{2}{c}{\bf CMB+BAO} &  \multicolumn{2}{c}{\bf CMB+BAO+lensing} \\
\noalign{\vskip 3pt}\cline{2-7}\noalign{\vskip 3pt}

 Parameter &  Best fit &  68\% limits &  Best fit &  68\% limits &  Best fit &  68\% limits \\
\hline
{\boldmath$\omega_b       $} & $0.022304                  $ & $0.02245\pm 0.00034\quad(+1.1 \sigma)$ & $0.022261                  $ & $0.02247\pm 0.00035\quad(+2.2 \sigma)$ & $0.022332                  $ & $0.02240^{+0.00030}_{-0.00034}\quad(+1.4 \sigma)$ \\
\ \ \ ($\Gamma=0$) & $0.022211                  $ & $0.02213\pm 0.00029        $ & $0.022035                  $ & $0.02192\pm 0.00026        $ & $0.022184                  $ & $0.02203\pm 0.00025        $ \\
\hline

{\boldmath$\omega_c       $} & $0.1368                    $ & $0.162\pm 0.023\quad(+16.0 \sigma)$ & $0.1392                    $ & $0.167^{+0.021}_{-0.018}\quad(+20.4 \sigma)$ & $0.1314                    $ & $0.1466^{+0.0092}_{-0.019}\quad(+15.0 \sigma)$ \\
\ \ \ ($\Gamma=0$) & $0.11911                   $ & $0.1195\pm 0.0026          $ & $0.12063                   $ & $0.1223\pm 0.0022          $ & $0.11864                   $ & $0.1201\pm 0.0018          $ \\
\hline

{\boldmath$H_0            $} & $83.6                      $ & $81^{+10}_{-10}\quad(-1.1 \sigma)$ & $69.30                     $ & $73.7^{+3.2}_{-5.8}\quad(+0.6 \sigma)$ & $69.05                     $ & $70.8^{+2.1}_{-3.6}\quad(-0.4 \sigma)$ \\
\ \ \ ($\Gamma=0$) & $98.6                      $ & $> 85.4                    $ & $69.42                     $ & $72.2^{+1.6}_{-2.9}        $ & $68.88                     $ & $71.6^{+1.4}_{-2.4}        $ \\
\hline

{\boldmath$\tau           $} & $0.0892                    $ & $0.089^{+0.012}_{-0.014}\quad(-0.0 \sigma)$ & $0.0870                    $ & $0.089^{+0.012}_{-0.014}\quad(+0.3 \sigma)$ & $0.0860                    $ & $0.081\pm 0.012\quad(+0.0 \sigma)$ \\
\ \ \ ($\Gamma=0$) & $0.0928                    $ & $0.089^{+0.012}_{-0.014}   $ & $0.0766                    $ & $0.084\pm 0.012            $ & $0.0860                    $ & $0.080^{+0.011}_{-0.012}   $ \\
\hline

{\boldmath$\Gamma/H_0     $} & $-0.120                    $ & $-0.29^{+0.22}_{-0.13}     $ & $-0.161                    $ & $-0.34^{+0.12}_{-0.14}     $ & $-0.121                    $ & $-0.232^{+0.14}_{-0.087}   $ \\
\ \ \ ($\Gamma=0$) & $0                         $ & $\mbox{---}                $ & $0                         $ & $\mbox{---}                $ & $0                         $ & $\mbox{---}                $ \\
\hline

{\boldmath$w_0            $} & $-1.53                     $ & whole prior range ($-3$,$-1$)  & $-1.100                    $ & $> -1.81\quad(-3.9 \sigma) $ & $-1.129                    $ & $> -1.34\quad(-1.9 \sigma) $ \\
\ \ \ ($\Gamma=0$) & $-1.939                    $ & $-1.52\pm 0.28             $ & $-1.035                    $ & $> -1.19                   $ & $-1.021                    $ & $> -1.14                   $ \\
\hline

{\boldmath$w_e            $} & $-2.18                     $ &  whole prior range ($-3$,$-1$)  & $-1.50                     $ & whole prior range ($-3$,$-1$) & $-1.043                    $ & $> -1.40\quad(-1.1 \sigma) $ \\
\ \ \ ($\Gamma=0$) & $-1.57                     $ & whole prior range ($-3$,$-1$)                         & $-1.148                    $ & $> -1.35                   $ & $-1.007                    $ & $> -1.21                   $ \\
\hline

{\boldmath$n_\mathrm{s}   $} & $0.9659                    $ & $0.9683\pm 0.0088\quad(+1.0 \sigma)$ & $0.9640                    $ & $0.9691\pm 0.0094\quad(+2.2 \sigma)$ & $0.9677                    $ & $0.9679^{+0.0071}_{-0.0087}\quad(+1.6 \sigma)$ \\
\ \ \ ($\Gamma=0$) & $0.9648                    $ & $0.9611\pm 0.0073          $ & $0.9594                    $ & $0.9550\pm 0.0064          $ & $0.9640                    $ & $0.9581\pm 0.0059          $ \\
\hline

{\boldmath$\ln(10^{10} A_\mathrm{s})$} & $3.0901                    $ & $3.082^{+0.024}_{-0.027}\quad(-0.2 \sigma)$ & $3.0836                    $ & $3.081^{+0.024}_{-0.027}\quad(-0.2 \sigma)$ & $3.0755                    $ & $3.061\pm 0.023\quad(-0.5 \sigma)$ \\
\ \ \ ($\Gamma=0$) & $3.0959                    $ & $3.088^{+0.024}_{-0.027}   $ & $3.0659                    $ & $3.085\pm 0.024            $ & $3.0799                    $ & $3.070^{+0.020}_{-0.023}   $ \\
\hline
$\Omega_{\mathrm{de}}      $ & $0.773                     $ & $0.695^{+0.15}_{-0.050}\quad(-3.7 \sigma)$ & $0.664                     $ & $0.647^{+0.058}_{-0.042}\quad(-4.4 \sigma)$ & $0.6775                    $ & $0.661^{+0.046}_{-0.031}\quad(-4.0 \sigma)$ \\
\ \ \ ($\Gamma=0$) & $0.8548                    $ & $0.817^{+0.043}_{-0.016}   $ & $0.7040                    $ & $0.723^{+0.015}_{-0.020}   $ & $0.7032                    $ & $0.722^{+0.013}_{-0.017}   $ \\
\hline

$\mathrm{Age}/\mathrm{Gyr} $ & $13.580                    $ & $13.72^{+0.15}_{-0.18}\quad(+2.7 \sigma)$ & $13.770                    $ & $13.792\pm 0.053\quad(+1.8 \sigma)$ & $13.790                    $ & $13.789\pm 0.054\quad(+1.6 \sigma)$\\
\ \ \ ($\Gamma=0$) & $13.423                    $ & $13.485^{+0.071}_{-0.10}   $ & $13.7474                   $ & $13.716\pm 0.043           $ & $13.7587                   $ & $13.720\pm 0.043           $\\
\hline
\hline
\end{tabular}
\caption{{\bf The phantom model ($\wde < -1$).} Best-fitting parameter values, mean and 68\% CL intervals for selected cosmological parameters (the primary MCMC parameters in bold, and derived parameters in non-bold face). For each parameter the first line is for the interacting dark-sector model and the second line for the non-interacting model ($\Gamma=0$). In parenthesis, the shift of the mean value compared to the non-interacting model is indicated in units of the standard deviation of the corresponding parameter of the non-interacting model.\label{bestfits_Phantom}}
\end{table}

\end{landscape}

\enlargethispage{0.3cm}
\bibliography{interactde2013}

\end{document}